\documentclass[graybox]{svmult}
\usepackage{mathptmx}       		
\usepackage{helvet}         			
\usepackage{courier}        			
\usepackage{type1cm}       	 	
\usepackage{makeidx}         		
\usepackage{graphicx}        		
\usepackage{multicol}        		
\usepackage[bottom]{footmisc}		
\makeindex             				

\begin{document}

\title*{PAMOP project:  computations in support of experiments and astrophysical applications}


\author{B M McLaughlin, C P Ballance, M S Pindzola, P C Stancil, S Schippers and \\ {A M\"{u}ller}}


\institute{B M McLaughlin	\at Centre for Theoretical Atomic, Molecular and Optical Physics (CTAMOP), 
					   School of Mathematics \& Physics, The David Bates Building,
       					   Queen's University, 7 College Park,  Belfast BT7 1NN, UK 
					\email{bmclaughlin899@btinternet.com}
\and C P Ballance \at 	     Centre for Theoretical Atomic, Molecular and Optical Physics (CTAMOP), 
					   School of Mathematics \& Physics, The David Bates Building,
       					   Queen's University, 7 College Park,  Belfast BT7 1NN, UK 
					\email{c.ballance@qub.ac.uk}
\and M S Pindzola \at 	     Department of Physics,  206 Allison Laboratory,
                            			     Auburn University, Auburn, AL 36849, USA 
					\email{pindzola@physics.auburn.edu}
\and P C Stancil   \at	Department of Physics and Astronomy and  the Center for Simulational Physics,
				University of Georgia, Athens, GA 30602-2451, USA
					\email{stancil@physast.uga.edu}
\and S Schippers  \at	I. Physikalisches Institut,
                         			     Justus-Liebig-Universit\"{a}t Giessen, 35392 Giessen, Germany 
					\email{Stefan.Schippers@physik.uni-giessen.de}
\and A M\"{u}ller  \at 	Institut f\"{u}r Atom- ~und Molek\"{u}lphysik,
                         			     Justus-Liebig-Universit\"{a}t Giessen, 35392 Giessen, Germany
					 \email{Alfred.Mueller@iamp.physik.uni-giessen.de}
					}


\maketitle

\abstract*{Our computation effort is primarily concentrated on support of current and future measurements 
		being carried out at various synchrotron radiation facilities around the globe,
		and photodissociation computations for astrophysical applications.
		In our work we solve the Schr\"odinger or Dirac equation 
		for the appropriate collision problem using the 
		R-matrix or R-matrix with pseudo-states approach from first principles.
		The time dependent close-coupling (TDCC) method is also used in our work.
		A brief summary of the methodology and ongoing developments 
		implemented in the R-matrix suite of Breit-Pauli and Dirac-Atomic R-matrix codes (DARC) is presented.} 		

\abstract{Our computation effort is primarily concentrated on support of current and future measurements 
		being carried out at various synchrotron radiation facilities around the globe,
		and photodissociation computations for astrophysical applications.
		In our work we solve the Schr\"odinger or Dirac equation 
		for the appropriate collision problem using the 
		R-matrix or R-matrix with pseudo-states approach from first principles.
		The time dependent close-coupling (TDCC) method is also used in our work.
		A brief summary of the methodology and ongoing developments 
		implemented in the R-matrix suite of Breit-Pauli and Dirac-Atomic R-matrix codes (DARC) is presented.} 		

\section{Introduction}
\label{subsec:0}
Our research efforts continue to focus on the development of computational 
methods to solve the Schr\"odinger and Dirac equations for atomic and 
molecular collision processes. Access to leadership-class computers such as the 
Cray XC40 at HLRS allows us to benchmark our theoretical solutions against dedicated collision 
experiments at synchrotron facilities such as the Advanced Light Source (ALS), 
Astrid II, BESSY II, SOLEIL and PETRA III and to provide atomic and molecular 
data for ongoing research in laboratory and astrophysical plasma science. 
In order to have direct comparisons with experiment, semi-relativistic, or 
fully relativistic computations, involving a large number of target-coupled 
states are required to achieve spectroscopic accuracy. These computations 
could not be even attempted without access to high performance computing (HPC) resources such 
as those available at leadership computational centers in Europe (HLRS) 
and the USA (NERSC, NICS and ORNL).
We use the R-matrix and R-matrix with pseudo-states (RMPS) methods 
to solve the Schr\"odinger and Dirac equations for atomic and molecular collision processes. 

Satellites such as {\it Chandra} and  {\it XMM-Newton}
are currently providing a wealth of x-ray spectra on many
astronomical objects, but a serious lack
of adequate atomic data, particularly in the {\it K}-shell energy range,
impedes the interpretation of these spectra.
 With the break-up and demise of the recently launched Astro-H satellite 
in the spring of 2016, it has left a void in x-ray  observational data 
 for a variety of atomic species of prominent astrophysical 
interest  of paramount importance (Kallman T,  Private communication, 2015).  
 In the intervening period before the next x-ray satellite mission,
we shall continue to benchmark laboratory photoionization cross section measurements 
against sophisticated theoretical methods.

The motivation for our work is multi-fold; (a) Astrophysical Applications 
\cite{McLaughlin2010,McLaughlin2001,Kallman2010,McLaughlin2013}, 
(b) Fusion and plasma modelling,  (c) Fundamental interest and (d) Support of 
experimental measurements and Satellite observations. 
For heavy atomic systems \cite{Ballance2012,McLaughlin2012},
little atomic data exists and our work provides results for new frontiers on the application of the 
R-matrix; Breit-Pauli and DARC parallel suite of codes.
Our highly efficient R-matrix codes are widely applicable to
the support of present experiments being performed at  synchrotron radiation facilities.   
Examples of our results  are presented below 
in order to illustrate the predictive nature 
of the methods employed  compared to experiment.

The main question asked of any method is, 
how do we deal with the many body problem? In our case
we use first principle methods (ab initio) to solve our dynamical equations of motion.  
Ab initio methods provide highly accurate, reliable atomic and 
molecular data (using state-of-the-art techniques) for solving the Schr\"{o}dinger  and Dirac equation.
The R-matrix non-perturbative method is used to model accurately a wide variety of 
atomic, molecular and optical processes such as; electron impact ionization (EII), 
electron impact excitation (EIE), single and double photoionization and inner-shell x-ray processes.
The R-matrix method provides cross sections and rates  used as input 
for astrophysical modeling codes such as; CLOUDY, CHIANTI, AtomDB, XSTAR 
necessary for interpreting experiment/satellite observations of astrophysical objects as well as 
fusion and plasma modeling for JET and ITER.  

\section{R-matrix code performance:  Photoionization}
\label{subsec:1}
The use of massively parallel architectures allows one to do calculations which previously could not have been addressed.
This  approach enables large scale relativistic calculations for trans-iron elements of Kr-ions, Xe-ions,
 Se-ions \cite{Ballance2012,McLaughlin2012} and W-ions \cite{Griffin2006,Griffin2013}. 
It allows one to provide atomic data in the absence of experiment, and for that 
purpose takes advantage of the linear algebra libraries available 
on most architectures. 
\begin{center}
\begin{table}
\caption{Photoionization cross section calculations:  
         timings for the $J=1$ 
         scattering symmetry of  W$^{2+}$ ions. 
         The scattering model used included 392-states, 
         1,728 coupled channels, and 800,000 energy points.
         The R-matrix outer region module PSTGBF0DAMP performance 
         on Hazel Hen, the Cray-XC40 at HLRS, is presented for a  
         different number of cores.}
\begin{tabular}{ccccc}
\hline
R-matrix		& Number of runs	& Speed-up 	& Cray XC40	& Total core time   \\
(Module)		&				& (factor)	& (Number of cores)	& (minutes)\\
\hline				
PSTGBF0DAMP		&	1			&  1.00				&	1,000		&	451.525\\
PSTGBF0DAMP		&	1			&  2.01				&	2,000		&	224.588\\
PSTGBF0DAMP		&	1			&  3.93				&	4,000		&	114.866\\
PSTGBF0DAMP		&	1			&  5.82				&	8,000		&	77.523\\
PSTGBF0DAMP		&	1			&  9.77				&   10,000	&	46.193\\
\hline
\end{tabular}
\end{table}
\end{center}
Further developments of the dipole codes benefit from similar 
 developments made to the existing excitation 
R-matrix codes \cite{McLaughlin2012,McLaughlin2015a,McLaughlin2015b,McLaughlin2016}.
In Table 1 we show typical timings required in the
determination of the photoionization cross section results for W$^{2+}$ ions, 
for the $J$=1 even scattering symmetry. Timings and speed up factors 
are given for the outer region module PSTGBF0DAMP used to determine 
photoionization cross sections. One clearly sees that using between 
1,000 to 10,000 cores, a speed up of nearly a factor 
of 10 is obtained with almost perfect scaling of this outer
region module.

\section{X-ray and Inner-Shell Processes}
%
%
%
\begin{figure}
\begin{center}
\includegraphics[scale=1,width=\textwidth]{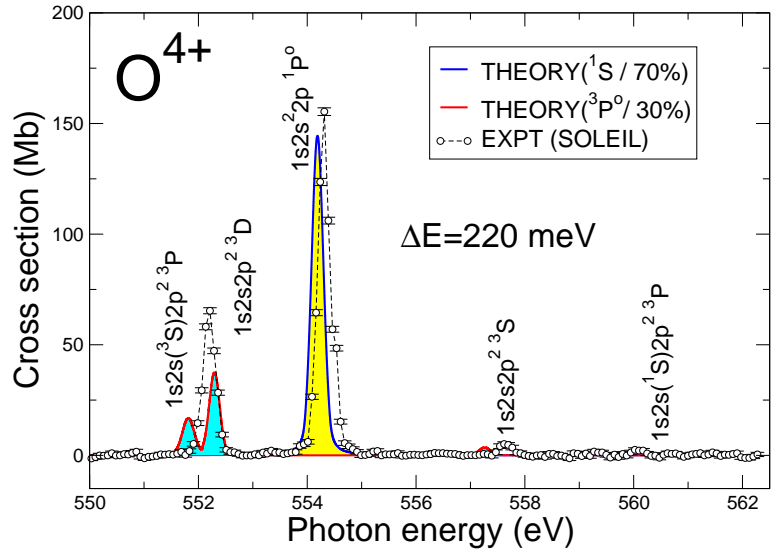}
\caption{\label{Figx1}  SOLEIL  experimental {\it K}-shell photoionization 
			cross section of O$^{4+}$ ions in the 550 - 560 eV photon energy range. 
		    Measurements were taken with a 220 meV band-pass at FWHM \cite{Soleil2016}. 
            Solid points (experiment): the error bars give 
			the statistical uncertainty.
             Solid line (R-matrix with pseudostates 526-levels) 
             assuming an admixture of
            70\% ($\rm 1s^22s^2\; ^1S$) and 30 \% ($\rm 1s^22s2p\; ^3P^o$). 
			The strong $\rm 1s \rightarrow 2p$ resonances are 
			clearly visible in the spectra.}
\end{center}
\end{figure}

%
%
%
\begin{figure}
\begin{center}
\includegraphics[scale=1,width=\textwidth]{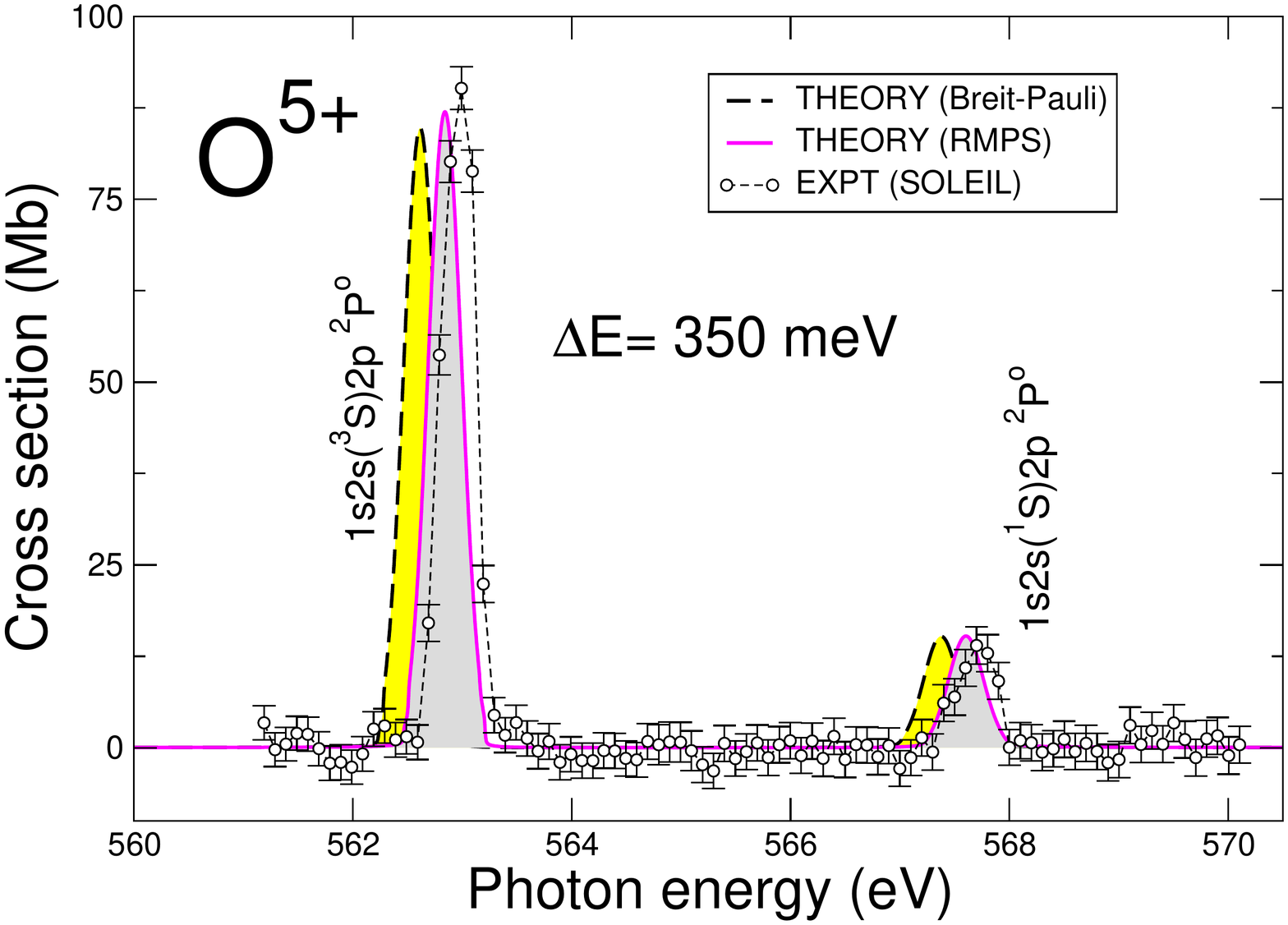}
\caption{\label{Figx2}  SOLEIL  experimental {\it K}-shell photoionization 
			cross section of O$^{5+}$ ions in the 560 - 570 eV photon energy range. 
		    Measurements were taken with a 350 meV band-pass at FWHM \cite{Soleil2016}.
            Solid points (experiment): the error bars give 
			the statistical uncertainty.
            Solid (magenta) line R-matrix with pseudostates, 
            120-levels for the $\rm 1s^22s\; ^2S$ ground state.
            Dashed (black) line Breit-Pauli approximation.
			The strong $\rm 1s \rightarrow 2p$ resonances are 
			clearly visible in the spectra.}
\end{center}
\end{figure}
 
\subsection{K-shell Photoionization of Atomic Oxygen Ions: O$^{4+}$ and O$^{5+}$}
\label{subsec:2}
The launch of the satellite Astro-H (re-named Hitomi) on February 17, 2016,
was expected to provide x-ray spectra of unprecedented quality and  
would have required a wealth of atomic and molecular data 
on a range of collision processes to assist with 
 the analysis of spectra from a variety of astrophysical objects.
 The subsequent break-up 40 days later on March 28, 2016 of Hitomi leaves 
 a void in observational x-ray spectroscopy. Measurements of  
cross sections for photoionization of atoms and ions are
essential data for testing theoretical methods in fundamental
atomic physics and for modeling of many physical
systems, for example, terrestrial plasmas, the upper atmosphere, 
and a broad range of astrophysical objects (quasar
stellar objects, the atmosphere of hot stars, proto-planetary
nebulae, H II regions, novae, and supernovae) 
\cite{Kjeldsen2002,Garcia2005}. 

Limited wavelength observations for x-ray transitions 
were recently made on atomic oxygen,
neon, magnesium and their ions with the 
High Energy Transmission Grating (HETG) 
on board the {\it CHANDRA}  satellite \cite{Liao2013}.  
Strong absorption {\it K}-shell lines
of atomic oxygen, in its various forms of 
ionization, have been observed by the XMM-Newton satellite 
in the interstellar medium, through x-ray 
spectroscopy of low-mass x-ray binaries \cite{Pinto2013}. 
The Chandra and XMM-Newton satellite 
observations may be used to identify absorption 
features in astrophysical sources, such as active galactic nuclei (AGN), 
x-ray binaries, and for assistance in  
benchmarking theoretical and experimental work 
\cite{Soleil2014b,Soleil2015,Gorczyca2013,Gatuzz2013a,Gatuzz2013b,Gatuzz2014}.

Absolute cross sections for the {\it K}-shell 
photoionization of  Be-like  (O$^{4+}$)
and Li-like (O$^{5+}$) atomic oxygen ions 
were measured (in their respective {\it K}-shell regions) by
employing the ion-photon merged-beam technique 
at the SOLEIL synchrotron-radiation facility in 
Saint-Aubin, France.  High-resolution spectroscopy with 
E/$\Delta$E $\approx$ 4000 ($\approx$ 140 meV, FWHM) 
 was achieved with photon energy from 550 eV up to 675 eV.
 Rich resonance structure observed  
 in the experimental spectra is analyzed using the 
 R-matrix with pseudosates (RMPS) method.

 Detailed spectra for
 Be-like [O$^{4+}$] and and Li-like [O$^{5+}$] atomic oxygen ions 
in the vicinity of the {\it K}-edge were measured.
 This work is the culmination of photoionization cross section 
measurements on the atomic oxygen isonuclear sequence.  
Previous  studies on this sequence, focused on obtaining photoionization 
cross sections for the O$^+$ and O$^{2+}$ ions \cite{Soleil2015} 
and the O$^{3+}$ ion \cite{Soleil2014b}, where differences of 0.5 eV 
in the positions of the $K_{\alpha}$ resonance  lines with prior satellite observations were found.  
This will have major implications for astrophysical modelling.

Fig. \ref{Figx1} shows the spectra for Be-like atomic oxygen in the region of the strong  
$\rm 1s \rightarrow 2p$ resonance.  To compare directly with the SOLEIL measurements,  
the theoretical R-matrix cross sections have been convoluted with a Gaussian  
profile width of 220 meV at FWHM.  For O$^{4+}$ as the $\rm 1s^22s2p\; ^3P^o$ metastable state is present 
in the photon beam, an admixture of 70\% of the ground state and 30\% of the metastable state, 
of the respective cross sections, appears to simulate experiment suitably well.  The theoretical cross section results
presented in Fig. \ref{Figx1} indicate excellent  agreement
with the SOLEIL experimental measurements. Similarly in
Fig. \ref{Figx2}, the SOLEIL spectra for Li-like atomic oxygen in the region of the strong  
$\rm 1s \rightarrow 2p$ resonance are illustrated.  To compare with the SOLEIL measurements,  
the theoretical cross sections have been convoluted with a Gaussian  
profile width of 350 meV at FWHM.
 We note that for both ions, the theoretical results from the R-matrix with pseudostates method (RMPS) 
show suitable agreement with the SOLEIL measurements \cite{Soleil2016}.

\subsection{{\it L}-shell Photoionization:  Ar$^{+}$}
\label{subsec:3}
%
%
%
%
\begin{figure}
\begin{center}
\includegraphics[scale=1.0,width=\textwidth]{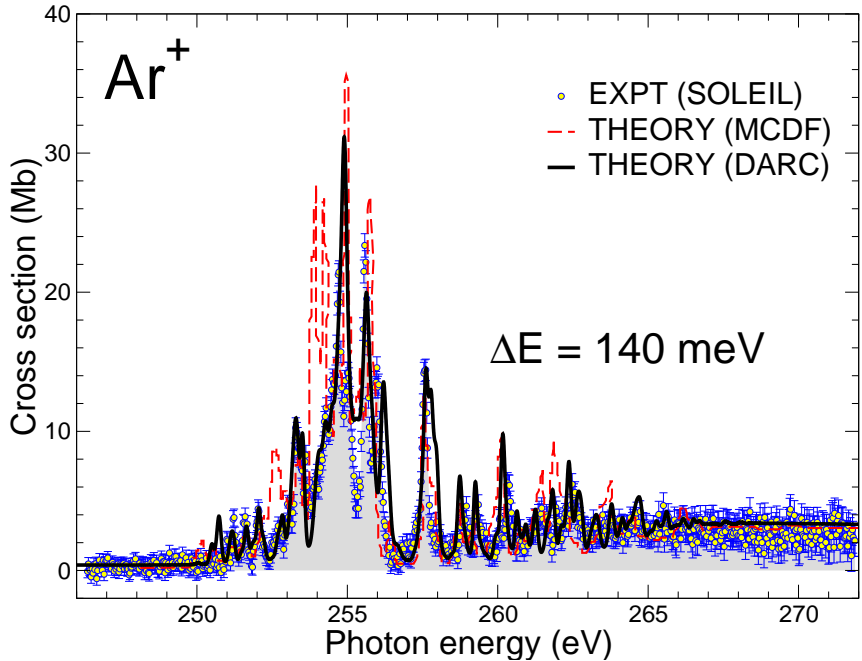}
\caption{\label{Ar+L}Photoionization cross sections (Mb) as a function of the 
	        photon energy (eV) in the Ar$^+$ {\it L}-shell region
                 between 250 and 270 eV. The (blue) circles
		are the experimental measurements from SOLEIL taken at a band pass of 140 meV at FWHM. 
		The dashed (red) line are the MCDF theoretical results and
		the solid (black) line are the DARC (model DARC3) results. The theoretical results 
                were statistically weighted for the initial ground state and 
		convoluted with a Gaussian profile width of 140 meV at FWHM \cite{Tyndall2016}.} 
\end{center}
\end{figure}

Photoionization cross-sections were obtained 
using the relativistic Dirac Atomic R-matrix Codes
(DARC) for valence and {\it L}-shell energy 
ranges between 27 and 270 eV. A total of 557
levels arising from the dominant configurations $\rm 3s^23p^4$, $\rm 3s3p^5$,
 $\rm 3p^6$, $\rm 3s^23p^3[3d, 4s, 4p]$, $\rm 3p^5 3d$,
$\rm 3s^2 3p^2 3d^2$, $\rm 3s3p^4 3d$, $\rm 3s3p^3 3d^2$,
 $\rm 2s^2 2p^5$ and $\rm 3s^2 3p^5$ have 
 been included in the target wavefunction
representation of the residual Ar$^{2+}$ion, 
including up to $\rm 4p$ in the orbital basis.
The target wavefunctions were obtained using 
the GRASP code \cite{dyall89,grant07}, and the collision 
calculations were performed using a parallel version of the DARC codes
 \cite{McLaughlin2015a,McLaughlin2015b,McLaughlin2016,darc}.
 Direct comparisons of the photoionization 
 cross sections in the valence region
 showed excellent agreement with previous R-matrix results and 
 ALS measurements \cite{Phaneuf2011}.

Photoionization cross section calculations were performed in the {\it L}-shell 
energy region between 250 and 280 eV  in order to compare directly with
the measurements made by  Bizau and co-workers 
at the SOLEIL radiation facility in France \cite{Blancard2012}. 
To compare directly with the SOLEIL measurements, 
theory was  convoluted with a 140 meV Gaussian
profile width at FWHM to match the experiment.

Fig. \ref{Ar+L} illustrates the photoionization cross-section, 
as a function of the incident photon energy in eV across the
{\it L}-shell threshold region from 250 to 270 eV. Comparisons are made
between the experimental results from SOLEIL, and theoretical work, MCDF and DARC. 
In order to match the SOLEIL experimental spectrum an energy shift of
7.5 eV to the DARC calculations was necessary \cite{Tyndall2016}.

\subsection{Photoionization of Tungsten (W) Ions: W$^{2+}$ and W$^{3+}$}
\label{subsec:4}
Although not directly relevant to fusion,  
photoionization of tungsten atoms and ions 
is interesting because it can provide 
details about spectroscopic aspects and, 
as time-reversed photorecombination, 
provides access to the understanding 
of one of the most important atomic collision 
processes in a fusion plasma, electron-ion 
recombination. R-matrix theory is a tool 
to obtain information about electron-ion 
and photon-ion interactions in general. 
Electron-impact ionization and recombination 
of tungsten ions have been  studied 
experimentally~
\cite{Mueller2015b,Rausch2011a,Stenke1995c,
Schippers2011b,Krantz2014,Spruck2014,Mueller2016b,Mueller2016c} 
while there are no detailed measurements on 
electron-impact excitation of tungsten atoms in any charge state. 
Thus, the present study on photoionization of 
these complex systems and comparison of 
the experimental data with R-matrix calculations 
provides benchmarks and guidance for 
future theoretical work on electron-impact excitation.

%
%
%
%

\begin{figure}
\begin{center}
\includegraphics[scale=0.5,width=12cm]{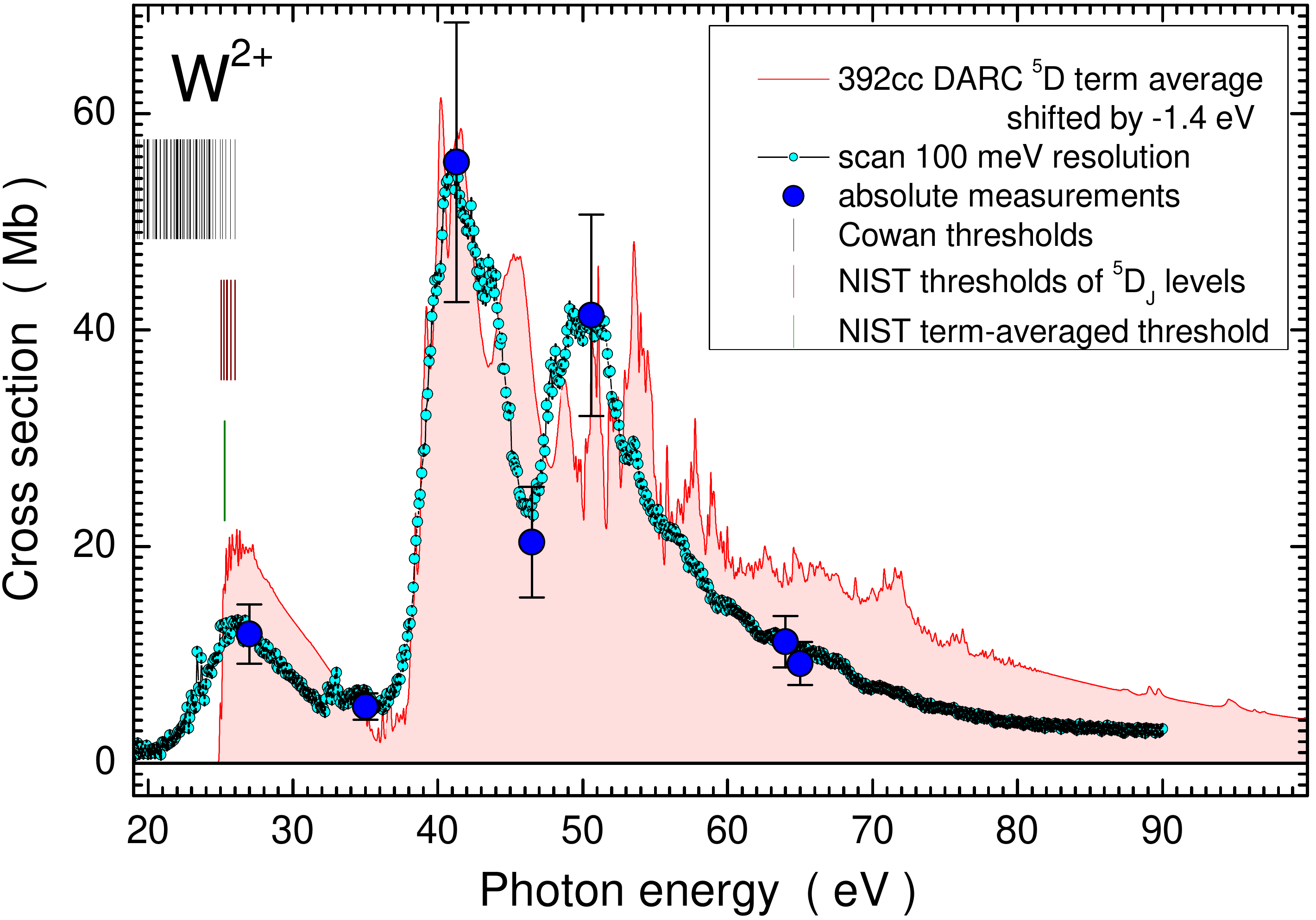}
\caption{\label{FigW2+} Photoionization of W$^{2+}$ ions  measured at energy resolution 100~meV.
 			Energy-scan measurements (small circles with statistical error bars) were normalized to
			absolute cross-section data represented by large circles with total error bars.	
			The black vertical bars at energies below 26~eV represent ionization thresholds
			of all $\rm 5d^4$, $\rm 5d^3 6s$, and $\rm 5d^2 6s^2$ levels with excitation
 			energies lower than the excitation energy of the lowest level (${\rm ^5G}_2$)
			within the  $\rm 5d^3 6p$ configuration. These thresholds were calculated by
			using the Cowan code~\cite{Cowan1981} as implemented
			by Fontes and co-workers~\cite{Fontes2015} and were shifted  by about 0.5~eV
			to match the ground level ionization threshold from the NIST tables~\cite{NIST2014}.
			The (brown) vertical bars between 25 and 26~eV indicate the NIST ionization
			potentials of the levels within the $\rm 5d^4\, ^5D$ ground-term.
			The lowest (green) vertical bar which matches the
			cross-section onset shows the NIST ground-term-averaged ionization potential.
			The solid (red) line with (light red) shading represents the result of the present
			392-level DARC calculation (125~$\mu$eV step size) of the
			ground-term-averaged photoionization cross section, convoluted with a
			Gaussian of 100~meV width.
			The theoretical cross sections are shifted by -1.4~eV
			to match  experiment \cite{Mueller2016a}.}
\end{center}
\end{figure}

%
%
%
%
\begin{figure}
\begin{center}
\includegraphics[scale=1.0,width=12cm]{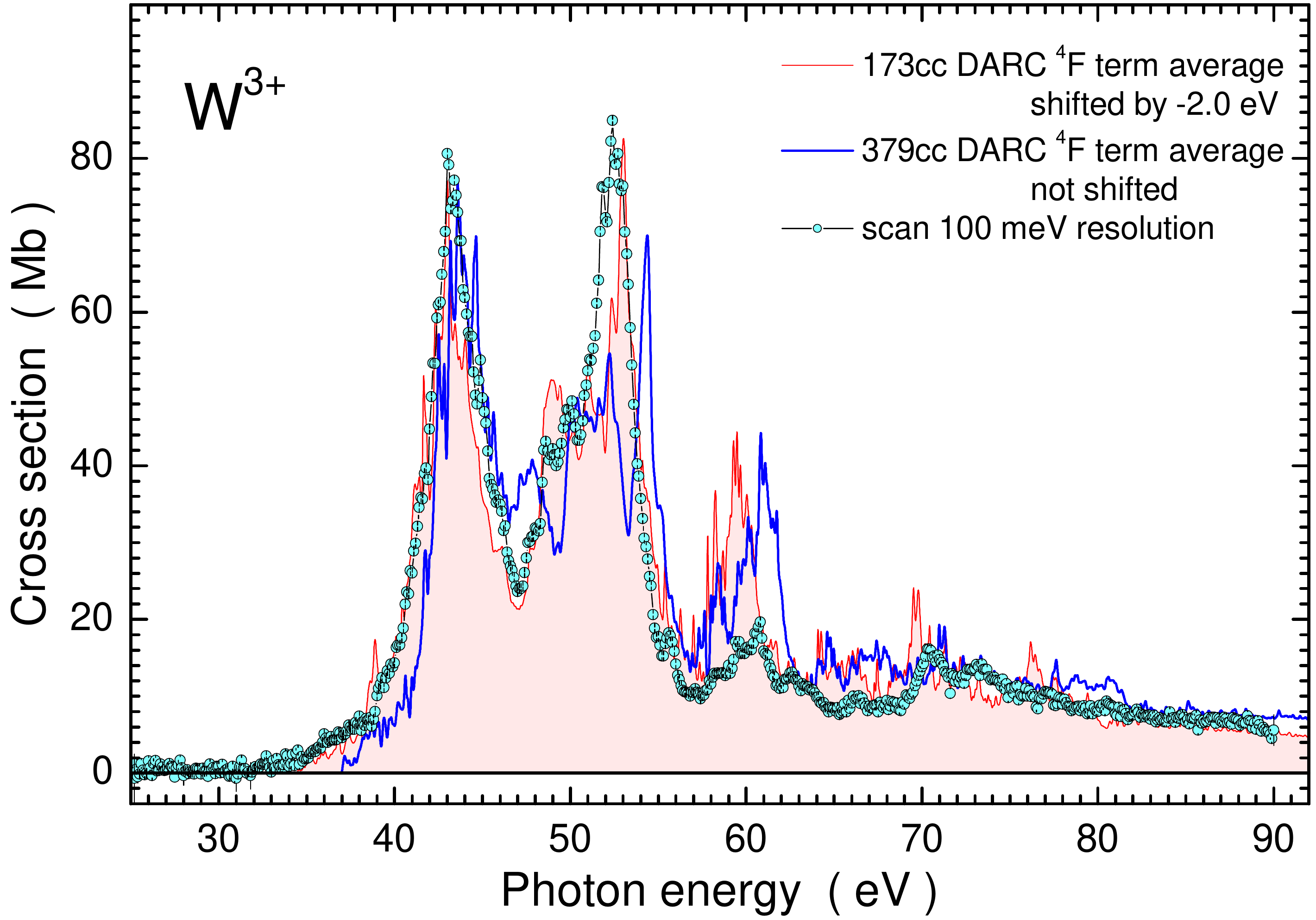}
\caption{\label{FigW3+} Comparison of the measured photoionization cross section of W$^{3+}$ with the
			present 173-level DARC calculation (87~$\mu$eV step size; thin red line with shading)
			and the present 379-level DARC result (109~$\mu$eV step size; solid blue line without shading).
			The theory curves were obtained by convolution of the
			original spectra with a Gaussian of 100~meV width.
			Only the 173-level calculations are shifted down in energy by 2.0~eV so that the steep rise of
			the experimental cross section function at about 40~eV is matched.}
\end{center}
\end{figure}

 For comparison with the measurements  made at the ALS, 
 state-of-the-art theoretical methods using
highly correlated wavefunctions were applied that include relativistic effects.
 An efficient parallel version \cite{Griffin2006,Griffin2013} 
 of the DARC \cite{norrington87,grant07,darc} suite
 of codes continues to be  developed
 and applied  to address electron and photon interactions with atomic systems,
 providing for hundreds of levels and thousands of scattering channels.
 These  codes are presently running on a variety of 
 parallel high performance computing
 architectures world wide \cite{McLaughlin2015a,McLaughlin2015b,McLaughlin2016}.
DARC calculations on photoionization of heavy ions carried out for   
Se$^{+}$ \cite{Ballance2012}, 
Xe$^{+}$ \cite{McLaughlin2012}, 
Fe$^+$ \cite{venessa2012}, 
Xe$^{7+}$ \cite{Mueller2014a},
W$^{+}$ \cite{Mueller2015c,Mueller2015d,Mueller2014c},
Se$^{2+}$ \cite{David2015}, and 
Kr$^{+}$\cite{Hino2012}, 
ions  showed suitable agreement with high resolution ALS measurements.
Large-scale DARC photoionization cross section calculations 
on neutral sulfur compared to photolysis experiments,  
made in Berlin \cite{Berlin2015}, and measurements performed at SOLEIL
for  2p removal in Si$^{+}$ ions by photons 
 \cite{Kennedy2014} both showed suitable agreement. 

Experimental and theoretical results are reported  for single-photon single ionization of W$^{2+}$ and W$^{3+}$  tungsten ions.  
Experiments were performed at the photon-ion merged-beam setup of the
Advanced Light Source in Berkeley. 
Absolute cross sections and detailed energy scans were measured over an
energy range from about 20~eV to 90~eV at a bandwidth of 100~meV. Broad peak features with  widths typically
around 5~eV have been observed with almost no narrow resonances present in the investigated energy range.
Theoretical results were obtained from a Dirac-Coulomb $R$-matrix approach. The calculations were carried
 out for the lowest-energy terms of the investigated tungsten ions with levels
${\rm 5s^2 5p^6 5d^4 \; {^5}D}_{J}$  $J=0,1,2,3,4$ for W$^{2+}$
 and  ${\rm 5s^2 5p^6 5d^3 \; {^4}F}_{J^{\prime}}$ $J^{\prime}=3/2, 5/2, 7/2, 9/2$ for W$^{3+}$.
 As illustrated in Fig. \ref{FigW2+} for W$^{2+}$ ions, suitable agreement is achieved below 60 eV, but at higher energies
there is a factor of approximately two difference between experiment and theory.
 In Fig. \ref{FigW3+}, assuming a statistically 
 weighted distribution of ions  in the initial ground-term levels, over the energy range investigated,
 good agreement between theory and experiment for W$^{3+}$ ions is achieved \cite{Mueller2016a}.  

\section{Single-Photon Double Ionization: He }
\label{subsec:5}
%
%
%
%
\begin{figure}
\includegraphics[width=\textwidth]{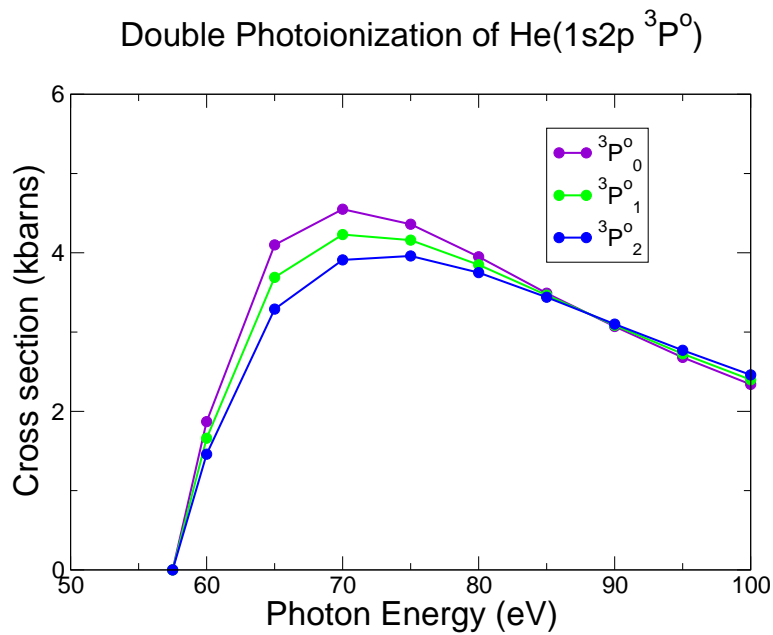}
\caption{\label{Fig1He}Total cross sections (kbarns) as a function 
			of photon energy using the time dependent 
			close-coupling (TDCC) method.  Results are shown 
	                for the initial individual fine-structure states
                    of $\rm He(1s2p~^3P^o_{J})$, where $J$=0, 1 and 2 \cite{Li2016}.}
\end{figure}
%
%
%
%
\begin{figure}
\includegraphics[width=\textwidth]{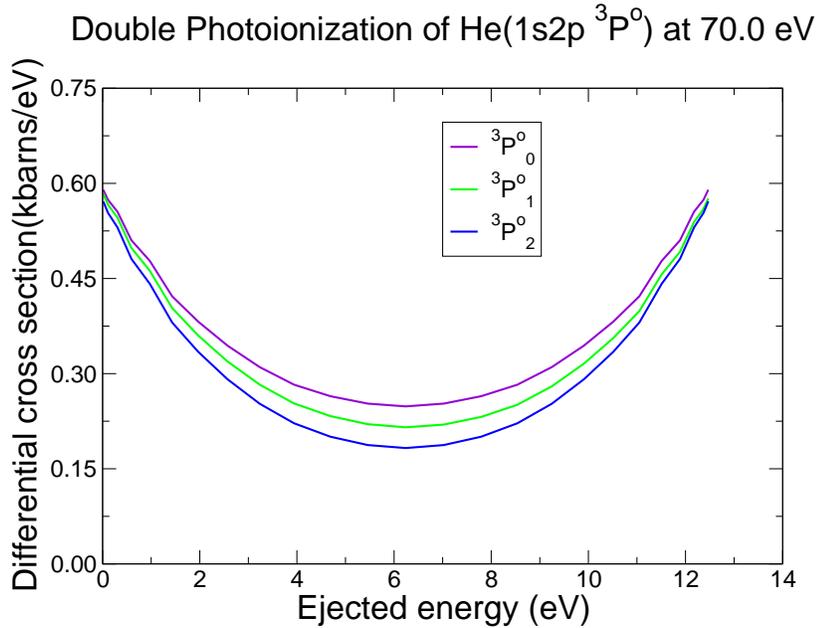}
\caption{\label{Fig2He}Differential cross sections 
               (kilobarns/eV) as a function of 
	           the ejected electron energy in eV using the 
               time dependent close-coupling (TDCC) method 
               at a photon energy of 70 eV. Results are shown for
               the initial individual fine-structure states 
	           of $\rm He(1s2p~ ^3P^o_{J})$, where $J$=0, 1 and 2 \cite{Li2016}.}
\end{figure}

The time-dependent close-coupling (TDCC) method 
\cite{Pindzola2007} was used to perform
single-photon double ionization cross section 
calculations of He in the $\rm 1s2p\; ^3P^o$ excited
state. Total and energy differential cross sections 
for the $\rm 1s2p~ ^3P^o$ excited state are presented for the 
TDCC ($\ell_1$, $\ell_2$, L) and
TDCC (${\ell}_1$ $j_1$,  ${\ell}_2$ $j_2$, J) representations.
Fig. \ref{Fig1He} illustrates the 
total TDCC total cross sections, and 
Fig. \ref{Fig2He} that for the differential cross section, 
as a function of the ejected electron energy in eV, for each initial 
 $\rm He(1s2p~ ^3P^o_{0,1,2})$ fine-structure level.
Differences found between the level resolved single-photon double
ionization cross sections are due to varying degrees
of continuum correlation found in 
the outgoing two electrons \cite{Li2016}. 

\section{Photodissociation: SH$^+$}
\label{subsec:6}
%
%
%
%
\begin{figure}
\includegraphics[scale=1.25,width=\textwidth]{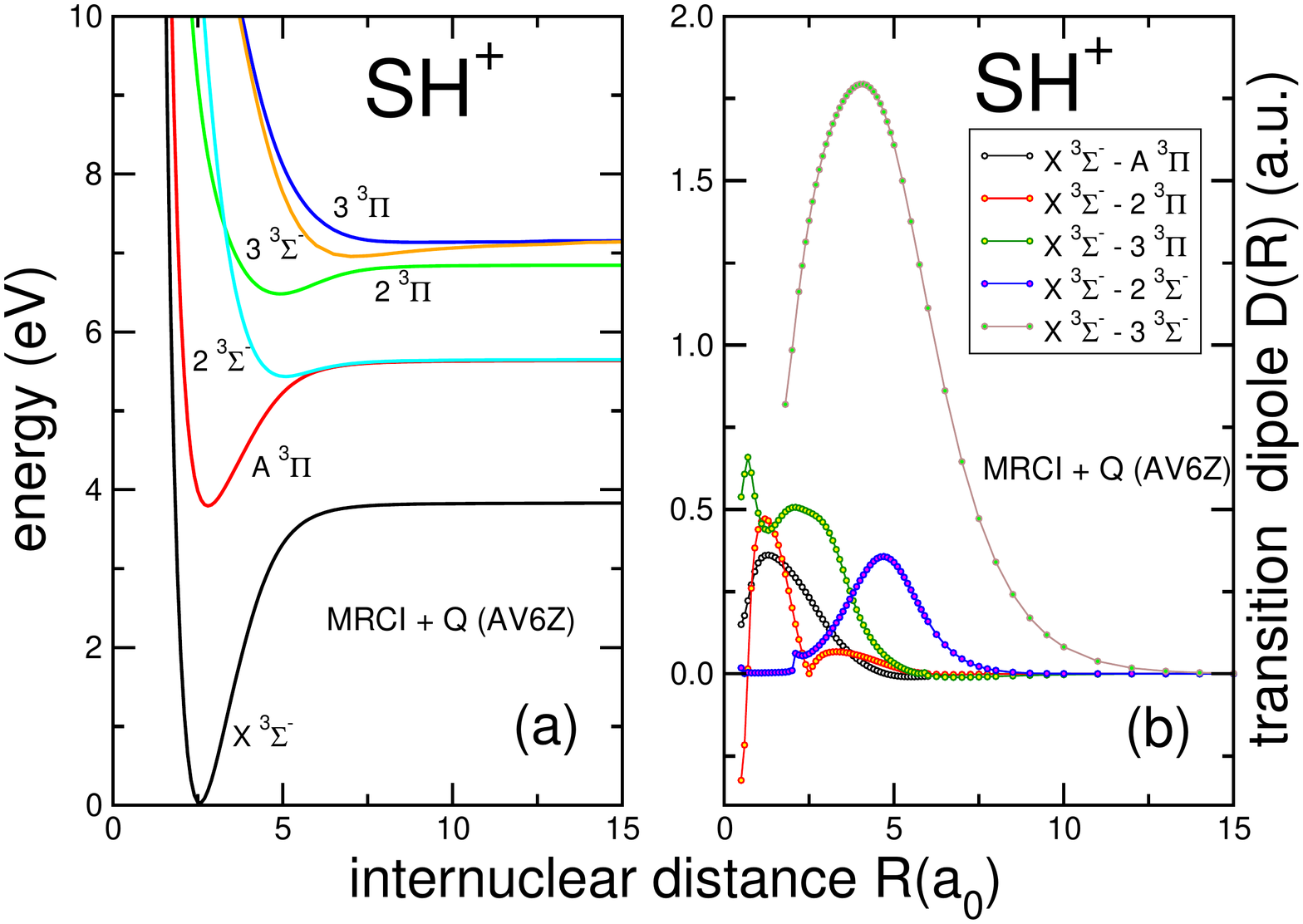}
\caption{\label{Fig1SH} (a) Relative electronic energies (eV) for the ${\rm SH}^+$ molecular cation, as a 
               function of bond separation at the MRCI+Q level of approximation with an AV6Z basis. 
               Energies are relative to the ground state near equilibrium (2.6 a$_0$).
               The states shown are for the transitions connecting the 
               X $^3\Sigma^{-} \rightarrow  2~^{3}\Sigma^{-}$, 
		3$~^{3}\Sigma^-$, A$~^3\Pi$, 2$~^{3}\Pi$, 3$~^{3}\Pi$ 
               states involved in the photodissociation process.
               (b) Dipole transition moments $D(R)$ (a.u.) for the ${\rm X}~^{3}\Sigma^{-} \rightarrow
 		{\rm A}~ ^{3}\Pi,  2~^{3}\Sigma^{-},   3~^{3}\Sigma^-, 2~^{3}\Pi, 2~^{3}\Pi, $ transitions.  The  MRCI + Q
		approximation with an AV6Z basis set  
         	 was used to calculate the transition dipole moments.}
\end{figure}

%
%
%
%
\begin{figure}
\includegraphics[scale=1.0,width=\textwidth]{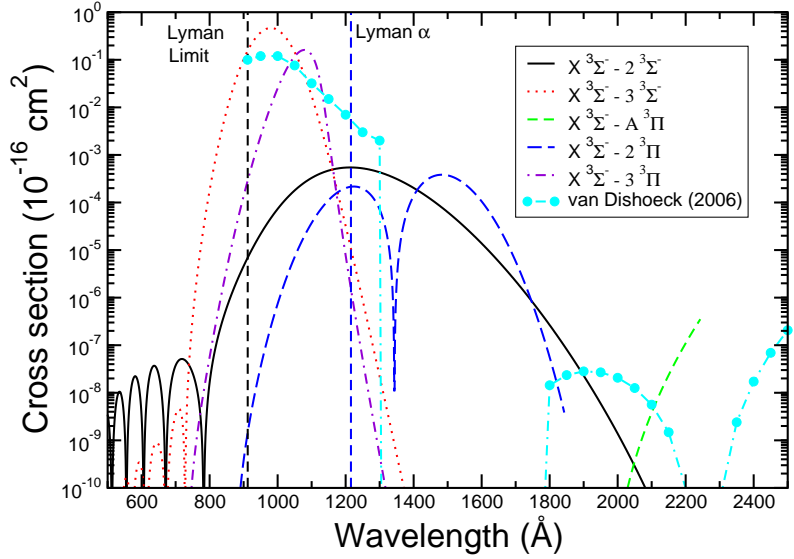}
\caption{\label{Fig2SH} Comparison of SH$^+$ photodissociation cross sections 
	 	for $v^{\prime\prime}=0$ and $J^{\prime\prime}=0$ 
		with estimates from Ref. \cite{Van2006}.}
\end{figure}
%
%
%
%
\begin{figure}
\includegraphics[scale=1.0,width=\textwidth]{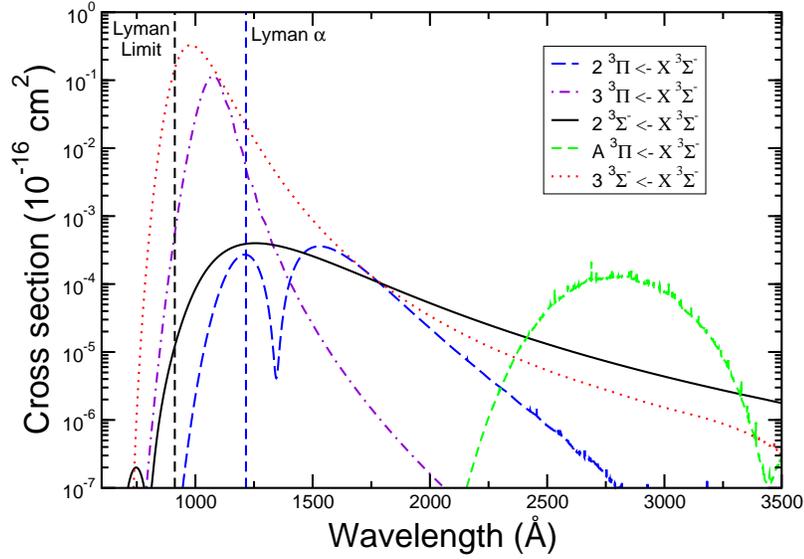}
\caption{\label{Fig3SH} Total SH$^{+}$ LTE photodissociation cross section at 
		3000 K for all electronic transitions.}
\end{figure}

Photodissociation cross sections for the SH$^+$ radical 
are computed from all rovibrational (RV)
levels of the ground electronic state X$~^3\Sigma^-$ 
for wavelengths from threshold to 500~\AA.
The five electronic transitions, $2~ ^3\Sigma^- \leftarrow$ X$~^3\Sigma^-$, 
$3~ ^3\Sigma^- \leftarrow$ X$~^3\Sigma^-$,
$A~ ^3\Pi \leftarrow$ X$~^3\Sigma^-$, $2~ ^3\Pi \leftarrow$ X$~^3\Sigma^-$, 
and $3~ ^3\Pi \leftarrow$ X$~^3\Sigma^-$,
are treated with a fully quantum-mechanical two-state model, 
(i.e. no non-adiabatic coupling between excited states 
was included in our work). The photodissociation calculations 
incorporate adiabatic potential energy curves (PEC) 
and transition dipole moment (TDM) functions   
computed in the multi-reference configuration interaction
 approach \cite{Helgaker2000} with the Davidson correction (MRCI+Q) \cite{Davidson1974}, 
 using an augmented-correlation-consistent polarized valence sextuplet 
basis set, designated as aug-cc-pV6Z or AV6Z, as illustrated in Fig. \ref{Fig1SH}. 
We have  adjusted our {\it ab initio} data to match 
available experimental molecular data and asymptotic
atomic limits. Local thermodynamic equilibrium (LTE) 
photodissociation cross sections were computed which
assume a Boltzmann distribution of RV levels in 
the X$~^3\Sigma^-$ molecular state of the SH$^+$ cation. 
The LTE cross sections are presented for 
temperatures in the range 1,000 - 10,000~K. 

As far as we are aware, the current work is the 
first explicit photodissociation calculations 
for the SH$^+$ radical ion. An estimate was made in 
van Dishoeck et al. \cite{Van2006} of the SH$^+$ cross section
by scaling that of CH$^+$. As illustrated in 
Fig.~\ref{Fig2SH}, there is suitable agreement, 
however the current results are about a 
factor of 3 larger, therefore we would expect 
the photodissociation rate
to be enhanced by a similar amount. 

In Fig.\ref{Fig3SH}, the LTE cross sections for 
all five transitions are compared at 3,000 K. 
This should be compared to Fig. \ref{Fig2SH} 
for  $v^{\prime\prime} =0$, $J^{\prime\prime} =0$
where it is seen that the cross sections are 
larger in the LTE case for wavelengths longer than $\sim$1500 \AA. 

The SH$^+$ radical ion, sulfanylium, was not 
detected in the interstellar medium (ISM) 
until as late as 2010 \cite{Ben2010}. 
It is however, an important tracer of gas 
condensations in dense regions and also probes
the warm surface layers of photo-dominated regions (PDR) \cite{Nagy2013}. 
Furthermore, its abundance is expected to be enhanced
 in x-ray dominated regions (XDR) \cite{Abel2008}. 
 In their model of the Orion Bar PDR,
Nagy et al. \cite{Nagy2013} find that 
photodissociation accounts for a maximum of about 4.4\% of
 the total destruction rate of SH$^+$, 
 since reactive collisions with H and dissociative 
 recombination by electrons are more efficient. 
 However, they adopted the estimated cross section 
of \cite{Van2006} for $v^{\prime\prime} =0$, $J^{\prime\prime} =0$. 
 We point out that the adoption of the current
cross sections would enhance the photodissociation 
contribution to greater than 10\%.
 We note that the photodissociation rates are 
 not given here as they are sensitive to
the local radiation field and dust properties. 
The latter is quite different in the Orion Bar 
from the average ISM of the galaxy. The densities 
and temperatures (10$^5$-10$^6$ cm$^{-3}$ and $\sim$1,000~K) 
of the Orion Bar PDR begin to approach the regime 
where photodissociation from 
excited states might contribute which is 
currently neglected in all models. 
Furthermore, LTE conditions are 
almost satisfied, but at 1,000~K there is 
not a significant difference 
between the LTE and 
$v^{\prime\prime} =0$, $J^{\prime\prime} =0$ 
cross sections \cite{McMillan2016}.

\section{Summary}
The power of the predictive nature of 
the R-matrix approach within a non-relativistic or a fully
relativistic approach for photoionization cross sections, 
valence or inner-shell, resonance energy positions, 
Auger widths and strengths is illustrated. 
Quantal calculation of photodissociation cross sections and rates 
for astrophysical applications require as input accurate potential 
energy curves and transition dipole moments.  
Access to leadership architectures is essential to our research work
such as the Cray-XC40 at HLRS which provides an integral 
contribution to our computational effort 
in atomic, molecular and optical collision processes.

\begin{acknowledgement}
A M\"{u}ller and S Schippers acknowledge support by 
Deutsche Forschungsgemeinschaft under project numbers Mu-1068/10, Mu-1068/20 
 and through NATO Collaborative Linkage grant 976362.  
B M McLaughlin acknowledges support from the 
US National Science Foundation through a grant to ITAMP
at the Harvard-Smithsonian Center for Astrophysics, under the visitor's program, 
the RTRA network {\it Triangle de le Physique} 
and a visiting research fellowship (VRF) 
from Queen's University Belfast. 
M S Pindzola acknowledges support by NSF 
and NASA grants through Auburn University.
P C Stancil acknowledge support by NASA grants
 through University of Goergia at Athens.
This research used computational resources 
at the National Energy Research Scientific 
Computing Center in Berkeley, CA, USA,  
and at the High Performance Computing 
Center Stuttgart (HLRS) of the University 
of Stuttgart, Stuttgart, Germany.
The Oak Ridge Leadership Computing 
Facility at the Oak Ridge National Laboratory, 
provided additional computational resources, 
which is supported by the Office of Science 
of the U.S. Department of Energy under Contract No. DE-AC05-00OR22725.
The Advanced Light Source is supported by the Director,
Office of Science, Office of Basic Energy Sciences,
of the US Department of Energy under Contract No. DE-AC02-05CH11231.
\end{acknowledgement}
%
%
%

\begin{thebibliography}{99.}%
%
%

%

\bibitem{McLaughlin2010} Hasoglu, M.~F., {A}bdel {N}aby, S. A.,  Gorczyca, T.~W., Drake J.~J., and McLaughlin, B.~M.:
					{\em K-shell Photoabsorption Studies of the Carbon Isonuclear Sequence.}
					Astrophys. J. \textbf{724}, 1296 (2010)
\bibitem{McLaughlin2001} McLaughlin, B. M.: {\em Inner-shell Photoionization, Fluorescence and Auger Yields.}
					In: Ferland, G. and Savin, D. W. (eds) {\em Spectroscopic Challenges of Photoionized Plasma}, 
					Astronomical Society of the Pacific, ASP Con$f$. Series \textbf{247} pp. 87. San Francisco (2001)

\bibitem{Kallman2010} Kallman, T. R.: {\em Challenges of Plasma Modelling: Current Status and Future Plansa.}
					Space Sci. Rev. \textbf{157}, 177 (2010)
\bibitem{McLaughlin2013} McLaughlin, B. M., and Ballance, C. P.:  {\em Photoionization, Fluorescence and Inner-shell Processes.}
    					 In: McGraw-Hill (eds) McGraw-Hill Yearbook of Science and Technology, pp. 281. Mc Graw Hill, New York (2013)

 \bibitem{Ballance2012} McLaughlin, B. M., and Ballance, C. P.: {\em Photoionization cross section calculations
				 	for the halogen-like ions Kr$^+$ and Xe$^+$.}  
					J. Phys. B: At. Mol. Opt. Phys. \textbf{45} 085701 (2012)
\bibitem{McLaughlin2012} McLaughlin, B., M., and Ballance, C. P.: 
					{\em Photoionization Cross-Sections for the trans-iron element Se$^{+}$ from 18 eV to 31 eV.}
		 			J. Phys. B: At. Mol. Opt. Phys. \textbf{45}, 095202 (2012)
\bibitem{McLaughlin2015a}{McLaughlin, B.~M., and Ballance, C.~P.:} 
					{\em Petascale computations for large-scale atomic and molecular collisions, Sustained Simulated Performance 2014} ed M M Resch,
					Y Kovalenko, E Fotch, W Bez and H Kobaysahi (New York: Springer) ch 15 (2014)

\bibitem{McLaughlin2015b}{McLaughlin, B.~M., Ballance, C.~P., Pindzola, M. S., and M\"uller, A.:} 
					{\em {PAMOP: petascale atomic, molecular and optical collisions:
					 High Performance Computing in Science and Engineering'14}}
					ed W E Nagel, D H Kr\"{o}ner and M M Resch (New York: Springer) ch 4 (2015)
					
\bibitem{McLaughlin2016}{McLaughlin, B.~M., Ballance, C.~P., Pindzola, M. S., Schipprs, S., and M\"uller, A.:} 
					{\em {PAMOP: petascale computations in suport of experiments: 
					High Performance Computing in Science and Engineering'15}}
					ed W E Nagel, D H Kr\"{o}ner and M M Resch (New York: Springer) ch 4 (2016)

 \bibitem{Griffin2006} Ballance, C. P.,  and Griffin D. C.: 
                        {\em Relativistic radiatively damped R-matrix calculation of the electron-impact excitation of W$^{46+}$.}
  				 J. Phys. B: At. Mol. Opt. Phys. \textbf{39}, 3617 (2006)

 \bibitem{Griffin2013} Ballance, C. P., Loch S. D., Pindzola M. S., and Griffin D. C.: {\em Electron-impact excitation 
 				 and ionization of W$^{3+}$ for the determination of tungsten influx in a fusion plasma.}  
					J. Phys. B: At. Mol. Opt. Phys. \textbf{46}, 055202 (2013)
					
\bibitem{Kjeldsen2002}Kjeldsen, H., Kristensen, B., Brooks, R. L., Folkman, H., Knudsen, H, and Andersen, T.:
	                      {\em Absolute state-selected measurements of the photoionization cross section of N$^+$ and O$^+$ ions.}
			      {Astrophys. J. Suppl. Ser.} {\textbf{138}} 219 (2002)


\bibitem{Garcia2005} Garcia, J., Mendoza, C., Bautista, M. A., Gorczyca, T. W., Kallman, T. R., and Palmeri P.:
				{\em K-shell Photoabsorption of Oxygen Ions.} 
				{Astrophys. J. Suppl. Ser.} \textbf{158} 68 (2005)				

\bibitem{Liao2013}Liao, J.-Y., Zhang, S.-N., and Yao, Y.:  
				{\em Wavelength Measurements of K Transitions of Oxygen, Neon, and Magnesium with X-ray Absorption Lines.}  
				{Astrophys. J.} {\textbf {774}} 116 (2013) 

\bibitem{Pinto2013}  Pinto, C., Kaastra, J. S., Costantini, E., and de Vries, C.:
				{\em Interstellar medium composition through X-ray spectroscopy of low-mass X-ray binaries.}
				{Astron. Astrophys.} {\textbf{551}} 25 (2013)

\bibitem{Soleil2014b} McLaughlin, B. M., Bizau, J. M., Cubaynes, D., Al Shorman, M. M., 
				Guilbaud, S., Sakho, I.,  Blancard, C. and Gharaibeh, M. F.:
				{\em K-shell photoionization of B-like ~(O$^{3+}$) oxygen ions: experiment and theory.}
				{J. Phys. B: At. Mol. Opt. Phys.} {\textbf{47}} 115201 (2014)
\bibitem{Soleil2015} Bizau, J. M., Cubaynes, D., Guilbaud, S., Al Shorman, M. M., Gharaibeh, M. F., 
				Ababneh, I. Q., Blancard, C. and McLaughlin, B. M.:
				{\em K-shell photoionization of O$^+$ and O$^{2+}$ ions: experiment and theory.}
				{Phys. Rev. A} {\textbf{92}} 023401 (2015)
\bibitem{Gorczyca2013}  Gorczyca, T. W., Bautista, M. A., Hasoglu, M. F., Garcia, J., Gatuzz, E., Kasstra, J. S., Kallman, T. R.,
                        Manson, S. T., Mendoza, C., Raasen, A. J. J., de Vries, C. P., and Zatsarinny, O.:
				{\em A comprehensive X-ray absorption model for atomic oxgen.}
				{Astrophys. J} {\textbf{779}} 78 (2013)
\bibitem{Gatuzz2013a}Gatuzz, E., Garcia, J. Mendoza, C., Kallman, T. R., Witthoeft,
				 M., Lohfink, A., Bauitista, M. A., Palmeri, P., and Quinet, P.:  
 		{\em Photoionization Modeling of Oxygen K Absorption in the Interstellar Medium: The Chandra Grating Spectra of XTE J1817-330.}`
				{Astrophys. J.}  {\textbf {768}} 60 (2013)

\bibitem{Gatuzz2013b}Gatuzz, E., Garcia, J. Mendoza, C., Kallman, T. R., Witthoeft,
				 M., Lohfink, A., Bauitista, M. A., Palmeri, P., and Quinet, P.:  
		{\em Erratum: Photoionization Modeling of Oxygen K Absorption in the Interstellar Medium: The Chandra Grating Spectra of XTE J1817-330.}`
				{Astrophys. J} {\textbf{778}} 83 (2013)

\bibitem{Gatuzz2014}  Gatuzz, E., Garcia, J., Mendoza, C., Kallman, T. R., Bautista, M. A., and Gorczyca, T. W:
                {\em Physical properties of the interstellar medium using high-resolution Chandra spectra: O K-edge absorption.}
				{Astrophys. J} {\textbf{790}} 131 (2014)

\bibitem{Soleil2016} McLaughlin, B. M., Bizau, J. M., Cubaynes, D., Guilbaud, S.,  Douix, S., Al Shorman,  M. M., El Ghazaly, M. O. A.,  
				 Sakho, I. and Gharaibeh, M. F.:
				{\em K-shell photoionization of O$^{4+}$ and O$^{5+}$ ions: experiment and theory.}
				{Mon. Not. Roy. Astro. Soc.} (MNRAS)  \textbf{465} 4690 (2017) 

\bibitem{dyall89}{Dyall, K. G., Grant, I. P., Johnson, C. T., and Plummer, E. P.:} 
	 			{\em GRASP: A general-purpose relativistic atomic structure program.}  
				{\em {Comput. Phys. Commun.}} {\bf \textbf{55}} 425 (1989)

\bibitem{grant07}{Grant, I., P.:}  {\em {Quantum Theory of Atoms and Molecules: Theory and
				  Computation.}} (New York, USA: Springer) (2007)

\bibitem{norrington87}{Norrington, P. H., and Grant, I. P.:}  
                 {\em Low-energy electron scattering by Fe XXIII and Fe VII using the Dirac R-matrix method.}
		 {{J. Phys. B: At. Mol. Opt. Phys.}} {\bf \textbf{20}} 4869 (1987)

\bibitem{darc}
				{R-matrix  DARC and BP codes, (2016):} \url{http://connorb.freeshell.org} 

\bibitem{Phaneuf2011} Covington, A. M., Aguilar, A., Covington, I. R., Hinojosa, G., Shirley, C. A., 
					Phaneuf, R. A., \'{A}lvarez, I., Cisneros, C., Dominguez-Lopez, I., 
					Sant'Anna, M.  M., Schlachter, A. S., Ballance, C. P. and McLaughlin, B. M.: 
					{\em Valence-shell photoionization of chlorinelike Ar$^{+}$ ions.}
					Phys. Rev. A  \textbf{84}, 013413 (2011)

\bibitem{Blancard2012}Blancard, C., Coss\'e, Ph., Faussurier, Bizau, J.-M., Cubaynes, D., El Hassan, N., Guilbaud, S.,
        	               Al Shorman, M. M., Robert, E., Liu, X.-J., Nicolas, C., and Miron, C.:
 	                      {\it L-shell} photoionization of Ar$^+$ to Ar$^{3+}$ ions.
				    {Phys. Rev. A} {\textbf{85}} 043408 (2012)
\bibitem{Tyndall2016}Tyndall, N. B., Ramsbottom, C. A., Ballance, C. P., and Hibbert, A.: {\em Valence and {\it L}-shell 
                                     photoionization of Cl-like argon using {\it R}-matrix techniques.}
			     Mon. Not. Roy. Astro. Soc. (MNRAS) {\textbf{456}} 366 (2016)

\bibitem{Mueller2015b}M\"{u}ller,  A.: {\em Fusion-Related Ionization and Recombination Data for Tungsten Ions in Low to
				Moderately High Charge States.}
				 {Atoms} {\textbf{3}} 120 (2015)
\bibitem{Rausch2011a}Rausch, J., Becker, A., Spruck, K., Hellhund, J., Borovik Jr, A.,
			      Huber, K., Schippers S., and M\"uller, A.: 
			     {\em Electron-impact single and double ionization of W$^{17+}$.}  
			     J. Phys. B: At. Mol. Opt. Phys. {\textbf{44}} 165202 (2011)
\bibitem{Stenke1995c}Stenke, M., Aichele, K., Harthiramani, D., Hofmann, G., 
				Steidl, M., V\"{o}lpel, R.,  and Salzborn E.: 
				{\em Electron-impact single-ionization of singly and multiply charged tungsten ions.} 
 				 {J. Phys. B: At. Mol. Opt. Phys.} {\bf 28} 2711 (1995)
\bibitem{Schippers2011b} Schippers, S., Bernhardt, D., M\"{u}ller, A., Krantz, C., Grieser, M., Repnow, R., Wolf, A.,
			  Lestinsky, M., Hahn, M., Novotn\'{y}, O. and Savin, D.~W.:
			  {\em Dielectronic recombination of xenonlike tungsten ions.} 
			 {Phys. Rev. A} {\bf 83} 012711 (2011)
\bibitem{Krantz2014}
			Krantz, C., Spruck, K., Badnell, N.~R., Becker, A., Bernhardt, D., Grieser, M., Hahn, M.,
			  Novotn\'{y}, O., Repnow, R., Savin, D.~W., Wolf, A., M\"{u}ller, A., and Schippers S.:  
			{\em Absolute rate coefficients for the recombination of open $f$-shell tungsten ions.}  
			  {J. Phys. Conf. Ser.} {\bf 488} 012051 (2014)

\bibitem{Spruck2014}
Spruck, K., Badnell, N.~R., Krantz, C., Novotn\'{y}, O., Becker, A., Bernhardt, D., Grieser, M.,
  Hahn, M., Repnow, R., Savin, D.~W., Wolf, A., M\"{u}ller, A., and Schippers S.: 
{\em Recombination of W$^{18+}$ ions with electrons: Absolute rate coefficients from a storage-ring experiment and 
from theoretical calculations.}  
   {Phys. Rev. A} {\bf 90} 032715 (2014)

\bibitem{Mueller2016b}{Borovik}, A. Jr., {Ebinger}, B., {Schury}, D., 
{Schippers}, S., and {M\"uller}. A.:
{\em Electron-impact single ionization of W$^{19+}$ ions.}
{Phys. Rev. A} {\bf 93} 012708 (2016)

\bibitem{Mueller2016c}{Badnell}, N. R., {Spruck}, K., {Krantz}, C., 
        {Novotn\'y}, O., {Becker}, A., {Bernhardt}, D., {Grieser}, M.,
        {Hahn}, M., {Repnow}, R., {Savin}, D. W., {Wolf}, A.,
         {M\"uller}, and {Schippers}, S.:
{\em Recombination of W$^{19+}$ ions with electrons: Absolute rate coefficients from a storage-ring experiment and from theoretical calculations.}
{Phys. Rev. A} {\bf 93} 052703 (2016)

\bibitem{venessa2012} Fivet, V., Bautista, M. A., and Ballance, C. P.:
				 {\em Fine-structure photoionization cross sections of Fe II.}
  				{J. Phys. B: At. Mol. Opt. Phys.} \textbf{45} 035201  (2012)

\bibitem{Mueller2014a}M\"{u}ller, A., Schippers, S., {Esteves-Macaluso}, D., Habibi, M., Aguilar, A., Kilcoyne, A.~L.~D.,
				 Phaneuf ,R.~A., Ballance, C.,~P., and McLaughlin, B.,~M.:  
  				{\em High Resolution Valence shell Photoionization of Ag-like (Xe$^{7+}$) Xenon ions : experiment and theory.}
				{J. Phys. B: At. Mol. Opt. Phys.} {\textbf 47} 215202 (2014)

\bibitem{Mueller2015c}M\"{u}ller, A., Schippers, S., Hellhund, J., Holosto, K., Kilcoyne, A. L. D., 
				Phaneuf, R. A., Ballance, C. P., and McLaughlin, B. M.:
				{\em Single-photon single ionization of W$^+$ ions: experiment and theory.}
				{J. Phys. B: At. Mol. Opt. Phys.} {\textbf{48}}  2352033 (2015)
\bibitem{Mueller2015d} M\"{u}ller, A.: {\em Precision studies of deep-inner-shell photoabsorption by atomic ions.} 
			 {Phys. Scr.} {\textbf {90}} 054004 (2015) 
\bibitem{David2015} Macaluso, D. A., Aguilar, A., Kilcoyne, A. L. D., Red, E. C., Bilodeau, R. C.,
                	    Phaneuf, R. A., Sterling, N. C., and McLaughlin, B. M.:
        	            {\em Absolute single-photoionization cross sections of Se$^{2+}$: experiment and theory.} 
               		     {Phys. Rev. A} {\bf 92} 063424 (2015)
\bibitem{Mueller2016a}McLaughlin, B. M., Ballance, C. P.,  Schippers, S., Hellhund, J., Kilcoyne, A. L. D., 
				Phaneuf, R. A.,  and  M\"{u}ller, A.:
				{\em Photoionization of tungsten ions: experiment and theory for W$^{2+}$ and W$^{3+}$.}
				{J. Phys. B: At. Mol. Opt. Phys.} {\textbf{49}} 065201 (2016)
\bibitem{Mueller2014c}M\"{u}ller, A., Schippers, S., Hellhund, J., Kilcoyne, A.~L.~D.,
				 Phaneuf, R.~A., Ballance, C.~P., and McLaughlin, B.~M.:
                                 {\em Single and multiple photoionization 
				of W$^{q+}$ tungsten ions in charged states $q$ = 1,2,..,5: experiment and theory.}
				 {J. Phys. Conf. Ser.} {\bf 488} 022032 (2014)
\bibitem{Cowan1981}Cowan, R. D.: {\em The Theory of Atomic Structure and Spectra.} 
                   {\em University of California Press, Berkeley, CA, USA} (1981)

\bibitem{Fontes2015}Fontes, C. J., Zhang, H. L., Abdallah, J. Jr., Clark, R. E. H., Kilcrease, D. P.,
                 Colgan, J. P., Cunningham, R. T.,  Hakel, P., Magee, N. H., and Sherrill M. E.: 
                 {\em The Los Alamos suite of relativistic atomic physics codes.}
                     {{J. Phys.  B: At. Mol. Opt. Phys.}} {\bf \textbf{48}} 144014 (2015)

\bibitem{NIST2014} {Kramida, A. E., Ralchenko, Y., Reader, J., and NIST ASD Team (2014),} {\em NIST Atomic
  Spectra Database (version 5.2),} National Institute of Standards and
  Technology, Gaithersburg, MD, USA

\bibitem{Hino2012}{Hinojosa, G., Covington, A. M., Alna'Washi, G. A., Lu, M., Phaneuf, R.~A., 
		Sant'Anna, M. M., Cisneros, C., {\'A}lvarez, I., Aguilar, A., Kilcoyne, A. L. D., 
			Schlachter, A.~S., Ballance, C. P., and McLaughlin, B.~M.:}
                              {\em Valence-shell single photoionization of Kr$^+$ ions: experiment and theory.}  
 			{{Phys. Rev. A}} {\bf 86} 063402 (2012)
\bibitem{Berlin2015}Barthel, M., Flesch, R., R\"{u}hl, E., and McLaughlin, B. M.: 
			{\em Photoionization of the  $3s^23p^4\;{\rm ^3P}$ and the $3s^23p^4\;{\rm ^1D, ^1S}$ states 
			of sulfur: experiment and theory.}
			{{Phys. Rev A}} \textbf{91} 013406 (2015)
\bibitem{Kennedy2014}Kennedy, E. T., Mosnier, J.-P., Van Kampen, P., Cubaynes, D., Guilbaud, S., Blancard, C.,
	          	McLaughlin, B. M., and Bizau, J.-M.:
	   	{\em Photoionization cross sections of the aluminumlike Si$^+$ ion in the region of the $2p$ threshold (94 -- 137 eV).} 
		{Phys. Rev. A} \textbf{90} 063409 (2014)
		
\bibitem{Pindzola2007}{Pindzola, M.S., Robicheaux, F., Loch, S. D., Berengut, J. C.,
                      Topcu, T., Colgan, J., Foster, M., Griffin, D. C.,
                      Ballance, C. P., Schultz, D. R., Minami, T., Badnell, N. R.,      
                      Witthoeft, M. C.,  Plante, D. R., Mitnik, D. M., Ludlow, J.A.,
                      and Kleiman, U.:}
                      {\em The time-dependent close-coupling method for 
                      atomic and molecular collision processes.}
				{J. Phys. B: At. Mol. Opt. Phys.} {\textbf{40}} R39 (2007)
\bibitem{Li2016}Li, Y., Pindzola, M. S., and Colgan, J. P.:
                 {\em Double photoionization of He from the $\rm 1s2p\,^3P^o$ excitated state.}
		 {{J. Phys. B: At. Mol. Opt. Phys.}} {\textbf{49}} 195205  (2016)

\bibitem{Helgaker2000} {Helgaker, T., Jorgesen, P., and Oslen, J.:} 
{\em Molecular Electronic-Structure Theory} (New York: Wiley) (2000)

\bibitem{Davidson1974} {Langhoff}, S. and {Davidson}, E. R.:
{\em Configuration interaction calculations on the nitrogen molecule.}
{Int. J. Quantum Chem.} {\bf 8} 61 (1974)

\bibitem{Van2006} van Dishoeck, E. F., Jonkheid, B., and van Hemert, M. C.:
                  {\em Photoprocesses in protoplanetary disks.} 
                  {Faraday Discuss.} \textbf{133} 855 (2006) 
\bibitem{Ben2010}Benz, A. O. 
{\it et al}: {\em Hydrides in young stellar objects: Radiation tracers in a protostar-disk-outflow system.} 
                  {Astron. Astrophys.} \textbf{521} A35 (2010) 

\bibitem{Nagy2013}Nagy, Z., {\it et al}: 
{\em The Chemistry of ions in the Orion Bar I. - CH$^+$, SH$^+$, and CF$^+$.}
                  {Astron.  Astrophys.} \textbf{550} A96 (2013) 
\bibitem{Abel2008}Abel, N.P.,  Federman, S. R., and Stancil, P. C.: 
{\em The  effects of doubly ionized
                  chemistry on SH$^+$ and S$^{2+}$ abundances in X-ray-dominated regions.}
                  {Astrophys. J} \textbf{675} L81 (2008) 
\bibitem{McMillan2016}McMillan, E. C., Shen, G., McCann, J. F., McLaughlin, B. M., and Stancil, P. C.: 
                     {\em Rovibrationally resolved photodissociation of SH$^+$.}
               		 {{J. Phys. B: At. Mol. Opt. Phys.}} {\textbf{49}} 084001  (2016)
\end{thebibliography}
%

\end{document}